\documentclass[a4paper, twocolumn]{article}
\usepackage[utf8]{inputenc}
\usepackage[hmarginratio=1:1,top=32mm,columnsep=20pt]{geometry}
\usepackage{amsmath}
\usepackage{amsfonts}
\usepackage{amssymb}
\usepackage{amsthm}
\usepackage{algpseudocode}
\usepackage{algorithm}
\usepackage[dvipsnames]{xcolor}
\usepackage{mathtools}
\usepackage[normalem]{ulem}
\usepackage{graphicx}
\usepackage{subcaption}
\usepackage{float}
\usepackage{commath}
\usepackage[noabbrev]{cleveref}
\usepackage{multirow}
\usepackage{caption}

\usepackage[
	backend=biber,
	style=numeric,
        maxbibnames=99,
        maxcitenames=1,
        uniquelist=false
	]{biblatex}
\renewbibmacro{in:}{%
  \ifentrytype{article}{}{\printtext{\bibstring{in}\intitlepunct}}}
\DeclareNameAlias{author}{last-first}
\theoremstyle{definition}

\theoremstyle{remark}

\newcommand{\N}{\mathbb{N}}

\usepackage{authblk}
\title{Topology of protein metastructure and $\beta$-sheet topology}
\author[1,2]{Jørgen Ellegaard Andersen}
\author[3,4]{Hiroyuki Fuji}
\author[2]{Yuki Koyanagi}
 \affil[1]{Danish Institute for Advanced Study, University of Southern Denmark}
\affil[2]{Centre for Quantum Mathematics, Department of Mathematics and Computer Science, University of Southern Denmark}
\affil[3]{Faculty of Information Science and Technology, Osaka Institute of Technology}
\affil[4]{Yukawa Institute for Theoretical Physics, Kyoto University}
\date{}
\begin{document}

\twocolumn[
\begin{@twocolumnfalse}
\maketitle  
\begin{abstract}
  \noindent
We introduce a new, simplified model of proteins, which we call protein metastructure. The metastructure of a protein carries information about its secondary structure and $\beta$-strand conformations. Furthermore, protein metastructure allows us to associate an object called a fatgraph to a protein, and a fatgraph in turn gives rise to a topological surface. It becomes thus possible to study the topological invariants associated to a protein.  We discuss the correspondence between protein metastructures and fatgraphs, and how one can compute topological invariants, such as genus and the number of boundary components, from fatgraphs. We then describe an algorithm for generating likely candidate metastructures using the information obtained from topology of protein fatgraphs. This algorithm is further developed to predict $\beta$-sheet topology of proteins, with a possibility to combine it with an existing algorithm. We demonstrate the algorithm on the data from PDB, and improve the performance of and existing algorithm by combining with it. 
\end{abstract}
\bigskip
\end{@twocolumnfalse}
]

\section{Introduction}
The configurations of $\beta$-strands in a protein, often called $\beta$-sheet topologies, have been studied since the 1970's \cite{richardson76}. $\beta$-sheets, along with $\alpha$-helices, are one of the fundamental structural components in the proteins. As opposed to helices, their structures involve interactions between residues which are far apart along the backbone. A better understanding of their structures and foldings is therefore crucial, if we are to understand the folding mechanism of entire proteins. The problem is further complicated by the intrinsic flexibility of $\beta$-sheet structures compared to $\alpha$-helices \cite{chothia97}.  Early studies \cites{richardson76, richardson77, sternberg77} have identified some general rules (such as the preference for the right-handedness in parallel $\beta$-sheets) from investigation of individual proteins. As the amount of available data increased, studies have used computer programs to survey the database and found frequent patterns in the $\beta$-strand configurations \cites{zhang00, ruczinski02}. The information can be used to filter and rank a series of decoy structures by computing probabilities for different patterns \cite{ruczinski02}. Another approach is to assign pseudoenergy to each pair of $\beta$-strand residues and solve the $\beta$-sheet topology prediction problem as an optimisation problem \cite{cheng05}. At least one study \cite{fonseca11} has compared the two methods, and found that the latter's performance to be better. One may also combine the two methods by, for example, forbidding certain $\beta$-strand configurations that are not found in the database \cite{subramani12}, or by incorporating the two in Bayesian modelling \cite{aydin11}. Other studies used integer programming techniques to predict $\beta$-sheet topologies \cites{savojardo13, eghdami15}. 

Fatgraph is a mathematical object, that has been used successfully to study topological structures of another biological macromolecule, RNA \cites{penner11, reidys11, andersen12}. A fatgraph can be thought of as a standard graph, where the edges and vertices have been ``fattened'' to ribbons and discs to form a surface (see \Cref{sec:fatgraph} for details). It has been particularly useful in solving the problem of enumerating topologically distinct RNA \cites{andersen13toprec, andersen13, andersen13enum} and protein structures \cite{andersen21enum}. Furthermore, the technique can be adopted to lower levels of abstraction by considering more parameters, thus allowing for enumeration of more realistic structures \cites{alexeev16, andersen16, andersen17p, andersen17enum}. It has also been shown that the topology of proteing fatgraphs is strongly linked to their geometric structures \cites{andersen21pre, andersen21using}. Inspired by their success, we introduce a new model for studying $\beta$-sheet topology of proteins, which we call protein metastructure. This model greatly simplifies the study of $\beta$-sheet topologies by amalgamating consecutive residues belonging to the same secondary structure, but still retains the information needed to understand the configuration of $\beta$-strands. We give a detailed definition in \Cref{sec:metastr}. Furthermore, each metastructure corresponds to a fatgraph, and this transition to fatgraphs allows us to compute topological invariants such as the number of boundary components and genus associated to each protein. The details of this correspondence are described in \Cref{sec:fatgraph}. Our use of fatgraphs in studying $\beta$-sheet topologies was inspired by \cite{penner10}, but our construction is much simpler, and can be constructed without the knowledge of proteins' geometric structures. We will use the topology of fatgraphs associated to proteins to predict $\beta$-strand conformations of proteins. More specifically, we will use the distribution of genus and number of boundary components to filter the candidate structures, whose topology does not agree with the distribution (see \Cref{sec:topochar} and \Cref{sec:metastr_apl} for details).

\subsection{Protein Metastructure}\label{sec:metastr}
\begin{figure}[h]
  \centering
  \begin{subfigure}[b]{.9\linewidth}
    \includegraphics[width=1\linewidth]{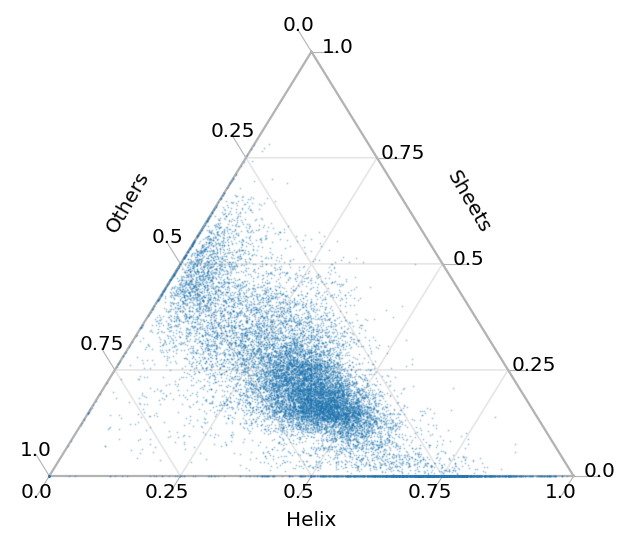}
    \caption{Proportion of 3 classes in secondary structure sequences}
    \label{fig:tri}
  \end{subfigure}
  
  \begin{subfigure}[b]{.9\linewidth}
    \includegraphics[width=1\linewidth]{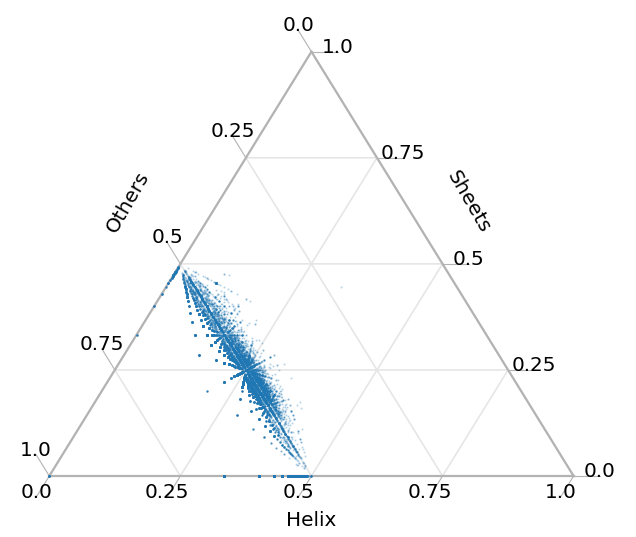}
    \caption{Proportion of 3 classes in reduced sequences}
    \label{fig:tri_reduced}
  \end{subfigure}
  \caption[DSSP class proportions]{Proportion of 3 classes in 16264 selected proteins}
  \label{fig:triangles}
\end{figure}

Given a protein, its primary structure is the sequence of amino acids in the polypeptide chain. There are 20 different amino acids in the standard gene code, so a primary structure can be expressed as a finite word in an alphabet with 20 letters;
\begin{equation}\label{pseq}
  \mathtt{EKKSINECDLKGKKVLIRVDFNVP...}
\end{equation}
The secondary structure of a protein can be defined as a set of local substructures, most frequent of which are $\alpha$-helices and $\beta$-sheets. The DSSP-algorithm \cite{kabsch83} is an algorithm commonly used to classify residues into 3 or 7 secondary structure classes. When used (with 3-class output) on the above protein it produces a word in  an alphabet with 3 letters;
\begin{equation}\label{secseq} \mathtt{\gamma\gamma\gamma\gamma\alpha\alpha\alpha\gamma\gamma\gamma\gamma\gamma\gamma\beta\beta\beta\beta\beta\beta\gamma\gamma\gamma\gamma\gamma...}
\end{equation}
Here we used the letter $\mathtt{\alpha}$ for Helices, $\mathtt{\beta}$ for Sheets, and $\mathtt{\gamma}$ for Coils. When we apply this reduction to the data extracted from PDB \cite{berman00} (see \Cref{sec:dataset} for details of data selection), we begin to see some patterns in the proportion of these classes in proteins. There are, for example, few proteins which contain less than 25\% $\mathtt{\gamma}$ residues, or more than 75\% of any one class (\Cref{fig:tri}). This can be explained by the rigidness of helix and sheet structures; a protein composed (almost) exclusively of $\alpha$ or $\beta$ residues will not have the necessary flexibility to bend and fold into its native structure. For that to occur, a certain proportion of $\gamma$ residues are required. On the other hand, too much $\gamma$ residues would most likely result in lack of stability and will be energetically unfavourable. The largest concentration appears to be around 30$\sim$50\% $\mathtt{ \alpha}$, 10$\sim$30\% $\mathtt{ \beta}$, and 30$\sim$50\% $\mathtt{ \gamma}$ residues (\Cref{fig:tri}).

We now introduce the \emph{reduced secondary structure sequence} by reducing each segment of identical letters in a secondary structure sequence (\ref{secseq}) to a single letter;
\begin{equation*}
  \mathtt{\gamma\alpha\gamma\beta\gamma\beta\gamma\beta\gamma\alpha...}
\end{equation*}
Not surprisingly, the distribution of proportions of the 3 classes in such reduced sequences are concentrated around $\mathtt{\gamma}=50\%$ (\Cref{fig:tri_reduced}), since the reduced sequences are mostly sequences of $\mathtt{\gamma\alpha}$s and $\mathtt{\gamma\beta}$s by construction.

In a reduced sequence, each letter $\beta$ corresponds to a $\beta$-strand. We may therefore add an additional data to a reduced sequence to specify $\beta$-sheet structure of the protein. To do this, we define the \emph{protein metastructure} as the triple $(r, P, A)$, where $r$ is a finite word in an alphabet of three letters, $\alpha, \beta$ and $\gamma$, and $P$ and $A$ are sets of pairs of integers $(i,j)$ for some $1 \leq i < j \leq s$, where $s$ is the number of letter $\beta$ in $r$. We also put a further condition, that $P \cap A = \emptyset$. Then for a given protein, we obtain its metastructure by setting $r$ to be the reduced sequence, and populating $P$ and $A$ as follows;

\begin{enumerate}
\item Number the letters $\beta$ in $r$ along the backbone, starting from the N-terminus.
\item Identify all pairs $(i,j)$, where there is at least one hydrogen bond between $i$th and $j$th strands.
\item Let $I$ be the set of all pairs $(i,j)$ identified in the previous step. Partition $I$ into two sets $P$ and $A$, where $P$ consists of all parallel connections and $A$ all anti-parallel connections.
\end{enumerate}
If there is only a single bond between two strands, thus making it impossible to determine the configuration between the two, we extend the strands by up to three residues. If it is still not possible to determine the configuration (because the extended strands has a single bond between them), then we assign the pair to $P$, as parallel configuration. This is because the standard anti-parallel configuration requires two hydrogen bonds between a pair of residues, thus making it less likely that there is only one hydrogen bond present. This forced assignment was necessary only in 181 out of 10141 proteins in the dataset, representing 1.8\% of the data.

\begin{figure}[h]
  \centering
  \includegraphics[width=.9\linewidth]{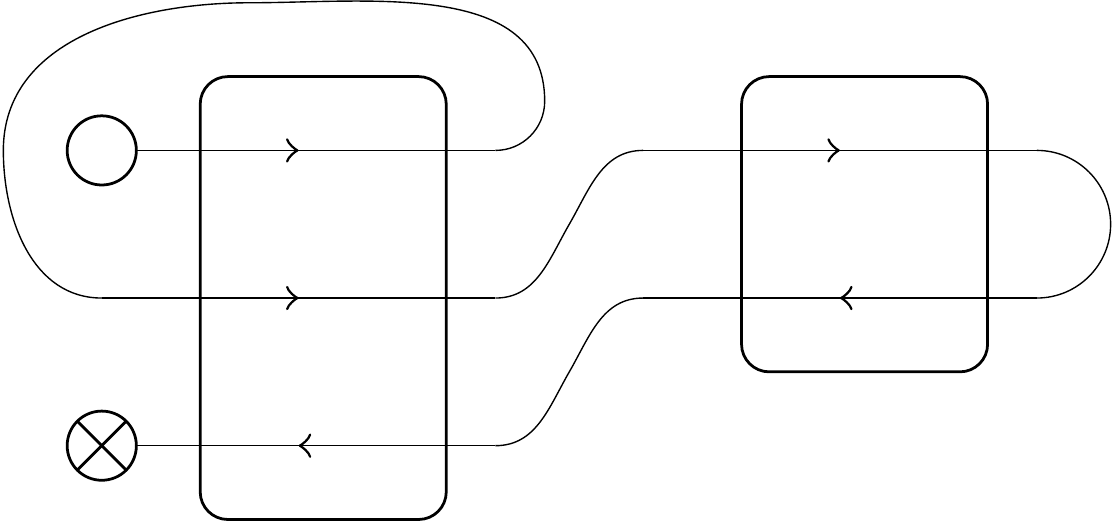}
  \caption[Metastructure motif diagram]{An example metastructure motif diagram. The associated metastructure may be $(\mathrm{\gamma\beta\gamma\alpha\gamma\beta\gamma\beta\gamma\beta\gamma\beta\gamma},
    \{(1,2)\}, \{(3,4), (2,5)\})$}
  \label{fig:motif}
\end{figure}

Let $\mathcal{S}$ be the set of all possible metastructures, and let $\mathcal{S}_{\mathrm{bif}} \subset \mathcal{S}$ be the subset consisting of all metastructures, where at least one $\beta$-strand is connected to more than 2 other strands (bifurcations).
Similarly, let $\mathcal{S}_{\mathrm{bar}}$ be the subset of metastructures with $\beta$-barrels (see \Cref{tab:filter} for the size of these subsets of metastructures). Consider $\tilde{\mathcal{S}} = \mathcal{S} \setminus \left( \mathcal{S}_{\mathrm{bif}} \cup \mathcal{S}_{\mathrm{bar}} \right)$. For each $s \in \tilde{\mathcal{S}}$, we can associate a metastructure motif diagram (\Cref{fig:motif}) as follows;

\begin{figure}[h]
  \centering
  \includegraphics[width=.9\linewidth]{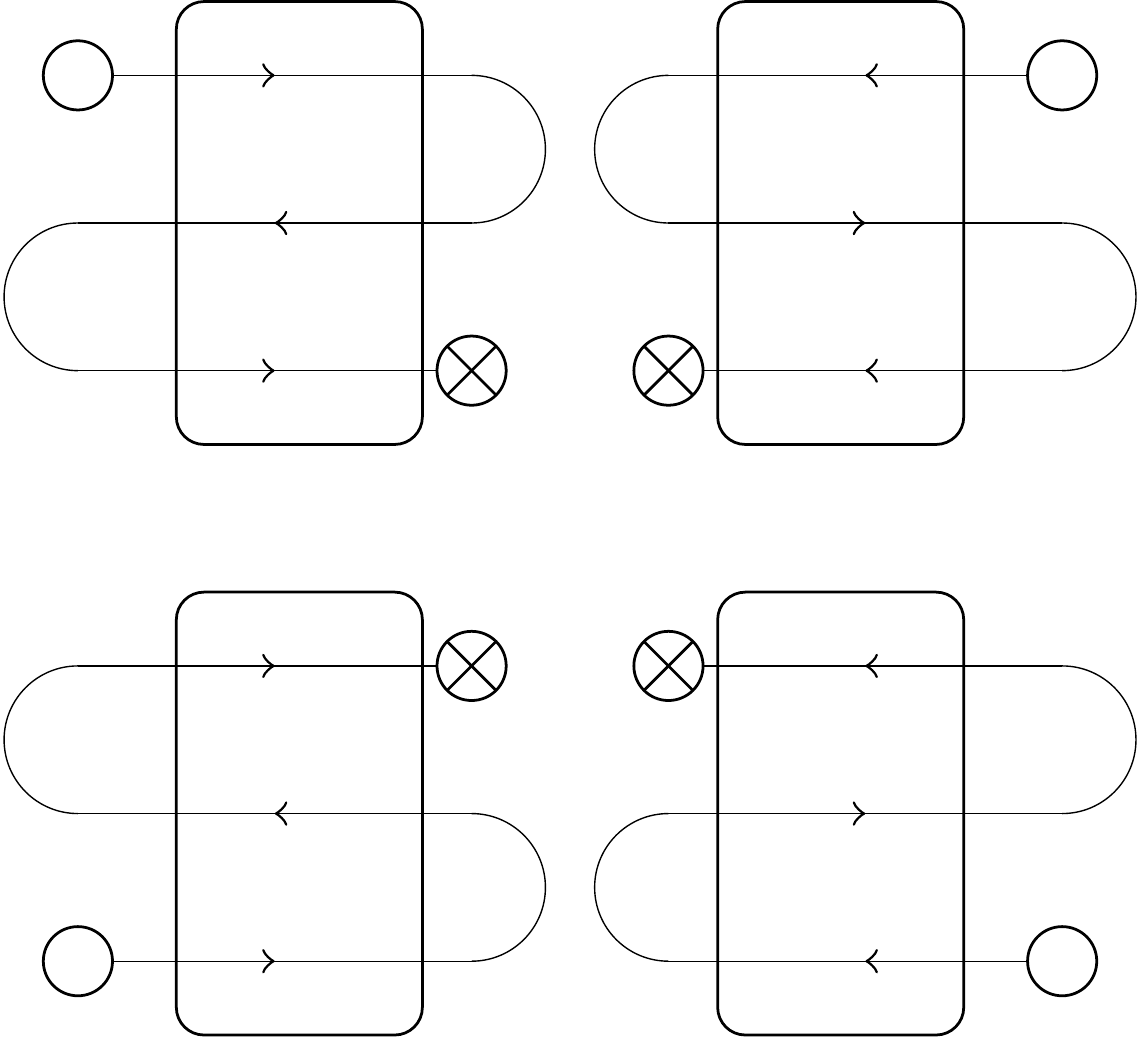}
  \caption[Equivalent metastructure diagrams]{Four equivalent metastructure diagrams. The two requirements force us to choose the top-left diagram.}
  \label{fig:metastr_choice}
\end{figure}

\begin{enumerate}
\item Each $\beta$-strand is denoted by a straight line segment with an arrowhead in the middle. 
\item If $(i,j) \in P$, draw the $i$'th and $j$'th strands next to each other, with arrowheads on both segments pointing the same direction.
\item If $(i', j') \in A$, draw the $i$'th and $j$'th strands next to each other, with arrowheads pointing the opposite direction.
\item Draw a ``sheet'' around each stack of strands.
\item Connect the strands, from the 1st to last, following the directions of arrowheads, and avoiding the interior of the sheets.
\item N-terminus is denoted by $\circ$, and C-terminus is denoted by $\scriptstyle{\otimes}$.
\item Note for each sheet, we have a choice of 2 strands to draw on the top, and a choice of which way the first strand points to (see \Cref{fig:metastr_choice}; orientation of all other strands are then decided by parallel/anti-parallel configurations). We can make this canonical by requiring that;
  \begin{enumerate}
  \item the top strand comes before the bottom strand in the backbone-ordering
  \item the first strand (in the backbone-ordering) in a sheet points from left to right
  \end{enumerate}
\end{enumerate}

\begin{figure}[h]
    \centering
    \includegraphics[width=.9\linewidth]{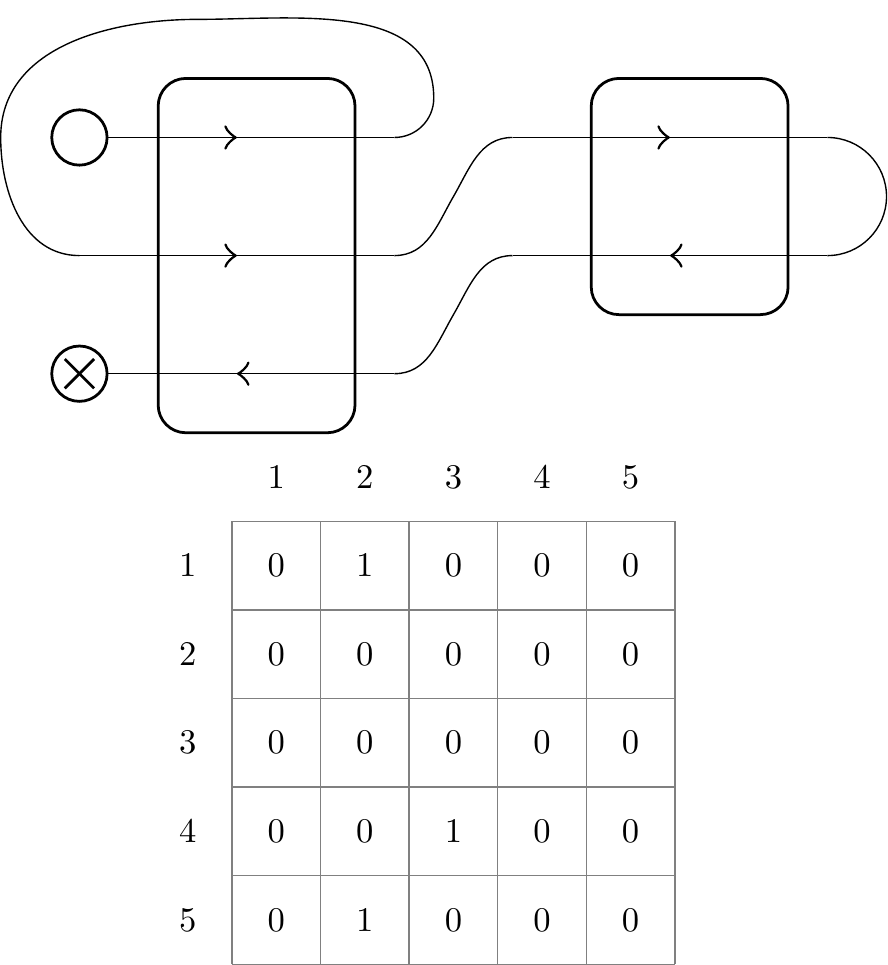}
    \caption[Pairing matrix]{A protein metastructure diagram and the corresponding pairing matrix. The 1 in (1,2)-th entry corresponds to the parallel configuration between the first and the second strand in the backbone, and the two 1's in the lower-triangular part correspond to the anti-parallel configurations between the third and the fourth strands, and the second and the fifth strands.}
    \label{fig:pairmat}
  \end{figure}

\begin{figure}[h]
    \centering
    \includegraphics[width=.9\linewidth]{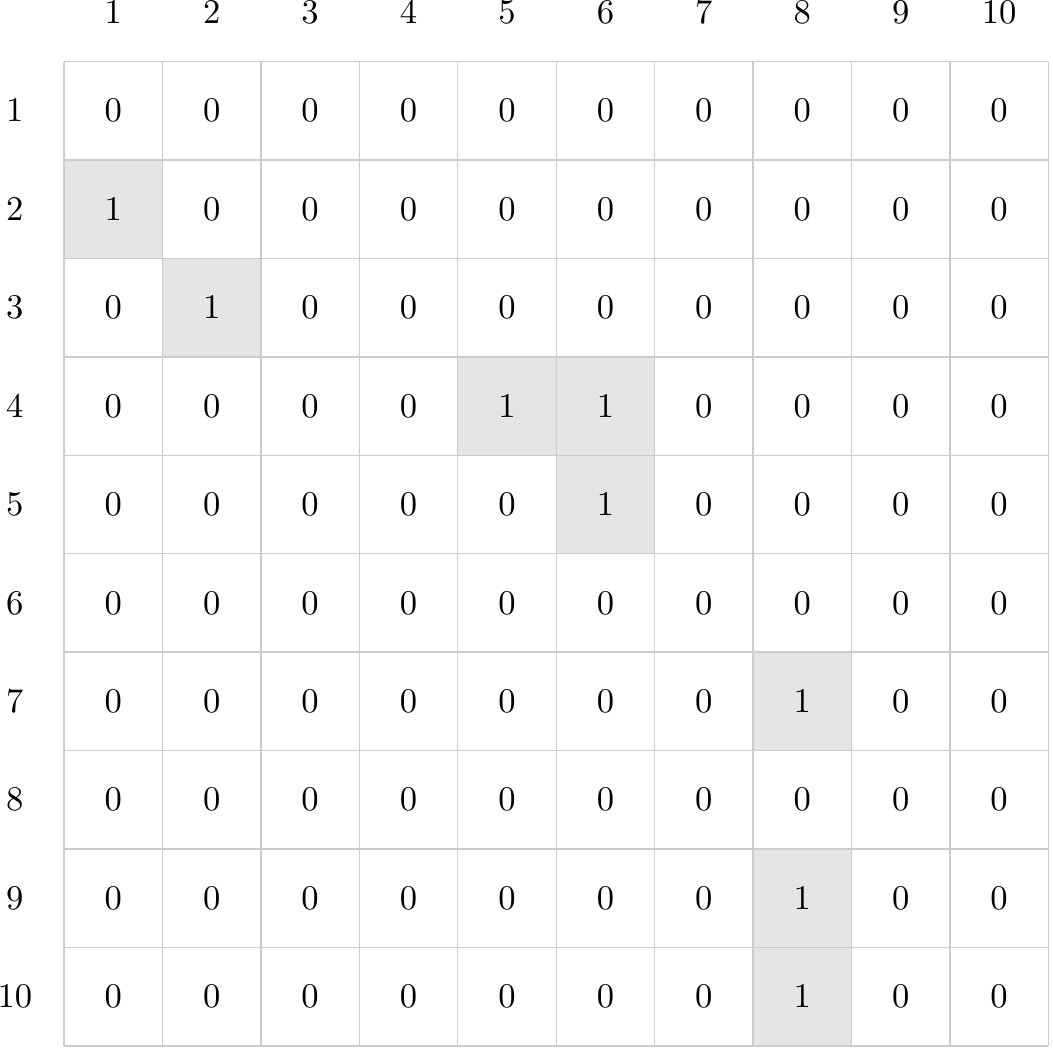}
    \caption[An example pairing matrix]{An example pairing matrix with three sheets. The first sheet consists of strands 1, 2, and 3 in anti-parallel configuration. The edge strands are 1 and 3, which can be seen by the fact that the total number of 1-cells in the first (or third) row and column is 1. On the other hand, the total number of 1-cells in the second row and column is 2, indicating the strand 2 is paired to two other strands. The second sheet, consisting of strands 4, 5, and 6, has no edge strand, and forms a barrel. The third sheet consists of strands 7, 8, 9, and 10. We see the strand 8 has three paired strands, thus indicating a bifurcation.}
    \label{fig:badpmat}
  \end{figure}

We note that the metastructure diagram, hence the $\beta$-sheet topology of a protein with $n$ $\beta$-strands can be recorded in an $n \times n$ matrix, whose entries are either 0 or 1. This can be done by setting $(i,j)$'th entry to 1 if the $i$'th and $j$'th strands are paired in the parallel configuration, and setting $(j,i)$'th entry to 1 if the pairing is anti-parallel. All other entries (where there is no pairing observed) are set to 0. We call this matrix $\mathbb{P}$ the protein's pairing matrix (\Cref{fig:pairmat}). The 1's in the upper-triangular part show parallel pairings, and the 1's in the lower-triangular part show anti-parallel pairings. The number of paired strands the $i$th strand has can be computed as the total number of 1 cells in the $i$th row and column. In a pairing matrix $\mathbb{P}$, an isolated strands can be seen as zero row and column; the $i$'th strand is isolated (has no paired strand), if and only if the $i$'th row and the $i$'th column do not have a 1. Similarly, the $i$'th strand has bifurcation, if and only if the total number of 1 cells in the $i$'th row and column is strictly greater than 2. A $\beta$-sheet manifests itself as a ``chain'' of strands, with the edge strand having only one non-zero entry in the corresponding row or column. A $\beta$-barrel is a circular chain without edge strands (\Cref{fig:badpmat}).

Note, if $\mathcal{T}$ is the set of metastructure motifs, the map $\varphi: \tilde{\mathcal{S}} \to \mathcal{T}$ described above corresponds to ``forgetting'' $r$ in $(r, P, A) \in \tilde{\mathcal{S}}$.

\subsection{Fatgraph}\label{sec:fatgraph}
\begin{figure}[h]
  \centering
  \includegraphics[width=.9\linewidth]{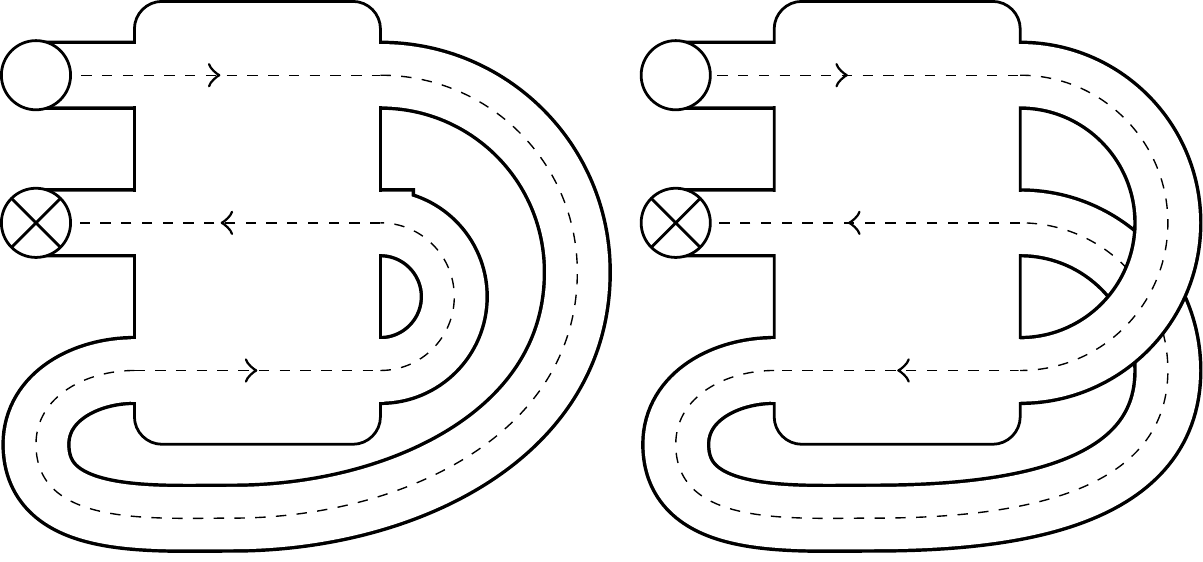}
  \caption[Thickening edges to obtain fatgraphs]{Thickening edges of metastructure diagrams to obtain fatgraphs (or more precisely, surfaces associated to fatgraphs). The surface on the left has genus 0, whereas the one on the right has genus 1.}
  \label{fig:fattening}
\end{figure}

\begin{figure}[h]
  \centering
  \includegraphics[width=.9\linewidth]{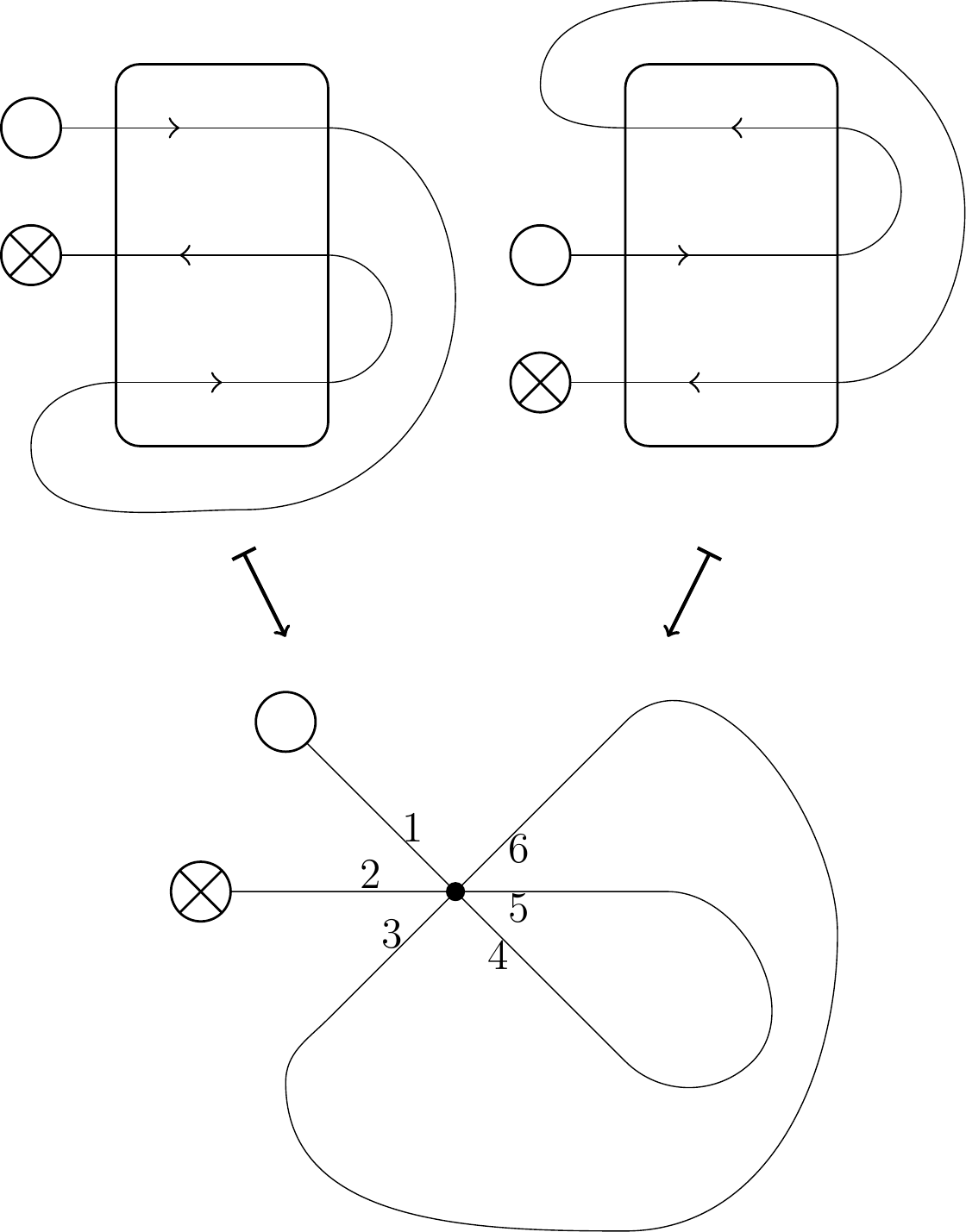}
  \caption[Fatgraph construction]{Construction of fatgraph from metastructure diagrams. Note the two different motifs result in an identical fatgraph.}
  \label{fig:fatgraph}
\end{figure}

In order to understand topological characteristics of protein metastructures, we need to pass from metastructure diagrams to topological surfaces. The main idea is to ``thicken'' the non-$\beta$ segments in a given metastructure diagram to (untwisted) bands or ribbons, as in \Cref{fig:fattening}, to produce a fatgraph $\mathbb{D}$.
Formally, a fatgraph $\mathbb{D}$ is a graph $D$ together with a cyclic ordering of the incident half-edges at each vertex. It can be obtained from a metastructure diagram by contracting each sheet to a point, and ordering the resulting half-edges at each vertex anti-clockwise from the N-terminus, or the starting end of the first strand in the sheet (\Cref{fig:fatgraph}). A fatgraph $\mathbb{D}$ gives rise to a unique (orientable) surface $X_{\mathbb{D}}$ by thickening each edge to a band and each vertex to a disc. As an orientable surface, it obeys Euler's formula
\begin{equation*}
  \chi (X_\mathbb{D}) = v - e + n = 2 - 2g,
\end{equation*}
where $v$ is the number of vertices (which correspond to the $\beta$-sheets in the metastructure diagram), $e$ the number of edges or bands (corresponding to the non-$\beta$ segments, excluding the N- and C-terminal segments), $n$ the number of boundary components, and $g$ the genus of $X_\mathbb{D}$.

Note this map $\psi$ from $\mathcal{T}$ to the set $\Sigma$ of fatgraphs with two marked half-edges is not injective (\Cref{fig:fatgraph}). Nonetheless the composition $\psi \circ \varphi$ allows us to compute topological invariants, such as genus and number of boundary components for protein metastructures.

\subsection{Topological characteristics of proteins}\label{sec:topochar}
\begin{figure*}[t]
  \centering
  \includegraphics[width=.9\linewidth]{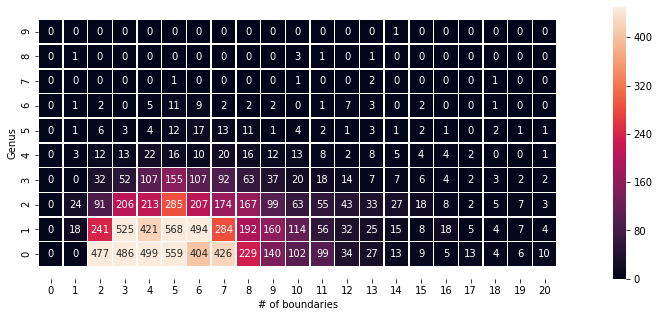}
  \caption[Frequency distribution, actual proteins]{Frequency distribution (extract) of protein metastructures by genus and number of boundary components}
  \label{fig:heatmap_pdb}
\end{figure*}

We compute genera and numbers of boundary components for 10141 selected proteins from PDB (\cite{berman00}; see \Cref{sec:dataset} for details of the selection process), which do not contain $\beta$-barrels or bifurcations in $\beta$-sheets. \Cref{fig:heatmap_pdb} shows frequency distribution of actual proteins by their genera and numbers of boundary components. 

\begin{figure}[!]
  \centering
  \includegraphics[width=\linewidth]{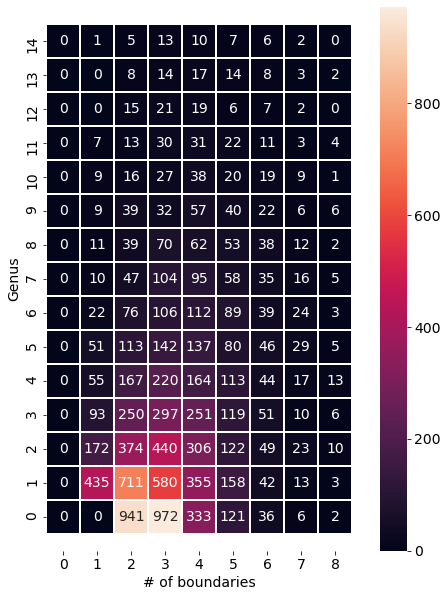}
  \caption[Frequency distribution, simulated proteins]{Frequency distribution (extract) of simulated protein metastructures by genus and number of boundary components}
  \label{fig:heatmap_sim}
\end{figure}
 
The same distribution was also computed from 10141 simulated metastructures, produced as follows;

\begin{enumerate}
\item Reduced sequences were generated in the following manner.
  \begin{enumerate}
  \item The length was chosen such that the distribution of lengths for the simulated data is the same as the distribution for the PDB data.
  \item Each pair of letters (1st and 2nd, 3rd and 4th, and so on) was given 50\% chance of being ``$\gamma\alpha$'' and 50\% chance of being ``$\gamma\beta$''. If the sequence has odd number of letters, the letter ``$\gamma$'' was attached at the end.
  \end{enumerate}

\item To each reduced sequence generated as above, a fatgraph structure was assigned as follows.
  \begin{enumerate}
  \item Let $U$ be the set of letter $\beta$s in a given sequence, indexed with their positions in the sequence; $\beta_1, \beta_2, \dotsc$. Then we partition $U$ into a random number of subsets, each containing at least 2 elements.
  \item For each subset $U_i$, choose a random ordering of $\beta_i$s in the subset. This defines the ordering of strands in a beta-sheet.
  \item For each ordered subset $U_i$ with $n_i$ elements, choose a random sequence of 1 and -1, of length $n_i$, but starting with 1. This sequence defines parallel/anti-parallel orientation of each strand with respect to the previous strand in a sheet.
\end{enumerate}
\end{enumerate}
We observe that the actual data tends to favour lower genera (and higher number of boundary components) compared to the simulated data (\Cref{fig:heatmap_sim}). This implies that metastructures whose associated surfaces have lower genera are favoured over those that result in high genera in the nature. Inspired by this observation, we will develop a method for prediction of $\beta$-sheet topology using the characteristics of the protein's associated surface in \Cref{sec:metastr_apl}.

For later use, we compute the distribution of the actual protein data by genus, number of boundary components and number of $\beta$-strands in the largest $\beta$-sheet. We call this the 3-dimensional genus-boundary distribution (see Supplementary Material).

  \section{Dataset}\label{sec:dataset}
  The dataset used was prepared similarly to the HQ60 dataset in \cite{penner14}. Here we give a brief summary of this dataset. PISCES \cite{wang03} is a service that, among other things, creates subsets of sequences from PDB based on specified threshold for structure quality and sequence identity. For the HQ60 dataset, we use only X-ray structures, with a resolution threshold of 2.0Å, Rfac threshold of 0.2, and maximum sequence homology of 60\%. The data was extracted from PDB in May 2021, resulting in a set of 16262 proteins. The hydrogen bonds are taken from the DSSP program \cite{kabsch83}, with the additional conditions \cite{baker84};
  \begin{align*}
    &\text{HO-distance} < 2.7\text{Å} \\
    &\text{angle(NHO)}, \text{angle(COH)} > 90^\circ.
  \end{align*}
  The secondary structures are also determined by DSSP, and they are recorded with three main secondary classes; [H]elix for H, G or I 8-state classes, [S]heet for E, and [C]oil for others. Thus we obtain, for each protein (of length $n$) in the dataset, a primary sequence $a_1a_2\cdots a_n$, where $a_i$ is one of the 20 standard gene code amino acids, and a secondary structure sequence $b_1b_2\cdots b_n$, where $b_i = \alpha, \beta, \text{ or } \gamma$. We superimpose these two sequences to obtain a \emph{hybrid sequence} $c_1c_2\cdots c_n$, where $c_i=b_i$ if $b_i=\alpha \text{ or } \beta$, and $c_i=a_i$ otherwise. Together with the information about hydrogen bonds, we are able to identify $\beta$-strands, their pairings and whether the pairing is parallel or anti-parallel (see \Cref{sec:metastr} for details). For the purpose of the current analysis, we are only interested in proteins containing $\beta$-sheets. Furthermore, proteins containing $\beta$-barrels are excluded from the analysis. For bifurcated $\beta$-sheets, we performed the following pre-processing;
  \begin{enumerate}
  \item For each $\beta$-strand $s$, let $p$ be the number of $\beta$-strands that are paired to it.
  \item If $p=3$, do the following.
  \item Let $s_1, s_2, s_3$ be the $\beta$-strands paired to $s$, ordered by the number of H-bonds to/from $s$. Let $n_1, n_2, n_3$ be the number of H-bonds between $s$ and $s_1, s_2, s_3$, respectively. We have $n_1 \geq n_2 \geq n_3$.
  \item If $n_3 \leq n_2/2$, ignore the pairing between $s$ and $s_3$.
  \end{enumerate}
  If, after the above pre-processing, a protein still contains a bifurcated $\beta$-sheet, it is excluded from the analysis. By performing the pre-processing, the number of proteins excluded because of bifurcation was reduced from 3859 to 1946. The procedure resulted in 11853 proteins for the analysis. See \Cref{tab:filter} for the number of proteins in each category. 

  \begin{table}[h]
    \centering
    \begin{tabular}{l r r}
      \hline
      $\alpha$ only  & 1551     & (9.5\%) \\
      Bifurcation    & 1946     & (12.0\%) \\
      $\beta$-barrel & 912      & (5.6\%) \\
      Accepted for analysis & 10141 & (72.9\%) \\
      \hline
      Total (HQ60)   & 16262    & \\
      \hline
    \end{tabular}
    \caption[Number of filtered proteins]{Number of proteins filtered from the dataset.}
    \label{tab:filter}
  \end{table}

\section{Methods}\label{sec:metastr_apl}
We will now describe a series of experiments to attempt to utilise the topological characteristics of protein metastructures described in \Cref{sec:topochar}.

\subsection{Binary classification of candidate structures by their topology}
\label{sec:binclas_metastr}
  200 proteins are randomly chosen for validation from the dataset, and the remaining 9941 proteins are used as the learning data. The idea is to use the learning data to decide the local configuration of $\beta$-strands, i.e. those strands, that are close to each other along the backbone. We then use the global topological data to decide the global configuration of the local blocks. We will now describe the first part of the method below. The aim is to first populate the pairing matrix $\mathbb{P}$ along the super- and sub-diagonals (i.e. the entries directly above and below the diagonal). We then repeat the procedure to populate the second entries above and below the diagonal, then the third entries, and so on. Pseudocode for populating the super- and sub-diagonals in the pairing matrix is given in \Cref{alg:binclas}.
  \begin{enumerate}
  \item From each protein in the validation data, we consider its hybrid sequence and extract segments between two consecutive $\beta$-strands.
  \item For each extracted segment $s$, compute alignment score for all segments from the learning data using the Needleman-Wunsch algorithm\footnote{For the substitution matrix we use blosum62 \cite{henikoff92}, extended by setting a match score with $\alpha$ or $\beta$ to 4 and mismatch involving $\alpha$ or $\beta$ to -4. See Supplementary Material for more details.} \cite{needleman70}.
  \item Let $t$ be the segment from the learning data with the highest alignment score. The configuration of two strands at either end of segment $t$ (whether they are paired by hydrogen bonds, and if so whether parallel or anti-parallel) determines the configuration of two strands at either end of $s$.
  \item Normalise the alignment score by dividing it by the score for perfect match, and record it in the appropriate entry in the pairing matrix $\mathbb{P}$. Specifically, if $p(s,t)$ is the alignment score for segments $s$ and $t$, the normalised score $\tilde{p}(s,t)$ is given by $p(s,t)/p(s,s)$. Suppose $s$ is the segment between $i$'th and $i+1$'th $\beta$-strands, and that they should be paired in the parallel configuration. Then set $\mathbb{P}_{(i,i+1)}=\tilde{p}(s,t)$. If, on the other hand, they should be paired in the anti-parallel configuration, set $\mathbb{P}_{(i+1,i)}=\tilde{p}(s,t)$
  \item If there is a tie for the highest alignment score such that $p(s,t_1) = p(s,t_2) = \cdots = p(s,t_k)$, set
    $x = \#\{t_i \vert \text{the two strands at either ends of } t_i \linebreak[1]\text{ are paired}\}$, where $\# S$ denotes the number of elements in a set $S$. Set $y=k-x$. The two strands at either end of $s$ are paired, if and only if $x \geq y$. The parallel/anti-parallel configuration of the two strands is determined similarly.
  \end{enumerate}

  \begin{algorithm}
    \begin{algorithmic}
      \State Let $t_1, t_2, \dotsc, t_m$ be the hybrid segments between two consecutive $\beta$-strands, extracted from all proteins in the learning dataset.
      \State Let $s_1, s_2, \dotsc, s_n$ be the hybrid segments between two consecutive $\beta$-strands in a given protein in the validation data.
      \State Let $\mathbb{P}$ by an empty $n \times n$ matrix.
      \For{$s_i$ in $s_1, s_2, \dotsc, s_n$}
      \For{$t_j$ in $t_1, t_2, \dotsc, t_m$}
      \State Compute alignment score $p(s_i, t_j)$.
      \EndFor
      \State Let $\tilde{j} \in \{1,2,\dotsc,m\}$ such that $p(s_i, t_{\tilde{j}}) = \max_j\{p(s_i, t_j)\}$.
      \If{$\tilde{j}$ is uniquely determined}
      \If{The two segments at either ends of $t_{\tilde{j}}$ are paired}
      \State Set $\tilde{p_i} = p(s_i, t_{\tilde{j}}) / p(s_i, s_i)$.
      \If{The two segments are paired in parallel configuration}
      \State Set $\mathbb{P}_{(i, i+1)} = \tilde{p_i}, \mathbb{P}_{(i+1, i)} = 0$.
      \Else
      \State Set $\mathbb{P}_{(i+1, i)} = \tilde{p_i}, \mathbb{P}_{(i, i+1)} = 0$.
      \EndIf
      \Else
      \State Set $\mathbb{P}_{(i, i+1)} = \mathbb{P}_{(i+1, i)} = 0$.
      \EndIf
      \Else
      \State Let $\tilde{j_1}, \tilde{j_2}, \dotsc, \tilde{j_k}$ be such that $p(s_i, t_{\tilde{j_h}}) = \max_j\{p(s_i, t_j)\}$ for all $h \in \{1,2,\dotsc,k\}$.
      \State Set $X = \{h | \text{two strands at either}\linebreak[1]\text{ ends of } t_{\tilde{j_h}} \text{ are paired}\}$
      \State Set $Y = \{1,2,\dotsc,k\} \setminus X$
      \If{$\# X \geq \# Y$}
      \State Set $\tilde{p_i} = p(s_i, t_{\tilde{j}}) / p(s_i, s_i)$.
      \State Let $P \subset X$ be the subset such that the two segments at either ends of $t_{\tilde{j_h}}, h \in P$ are paired in parallel configuration. 
      \State Let $A \subset X$ be the corresponding subset for anti-parallel configuration. 
      \If{$\# P \geq \# A$}
      \State Set $\mathbb{P}_{(i, i+1)} = \tilde{p_i}, \mathbb{P}_{(i+1, i)} = 0$.
      \Else
      \State Set $\mathbb{P}_{(i+1, i)} = \tilde{p_i}, \mathbb{P}_{(i, i+1)} = 0$.
      \EndIf
      \Else
      \State Set $\mathbb{P}_{(i, i+1)} = \mathbb{P}_{(i+1, i)} = 0$.
      \EndIf
      \EndIf
      \EndFor
    \end{algorithmic}
    \caption{Pseudocode for populating the first diagonal in the pairing matrix $\mathbb{P}$}
    \label{alg:binclas}
  \end{algorithm}
 
\begin{figure*}[t]
    \centering
    \includegraphics[width=.8\linewidth]{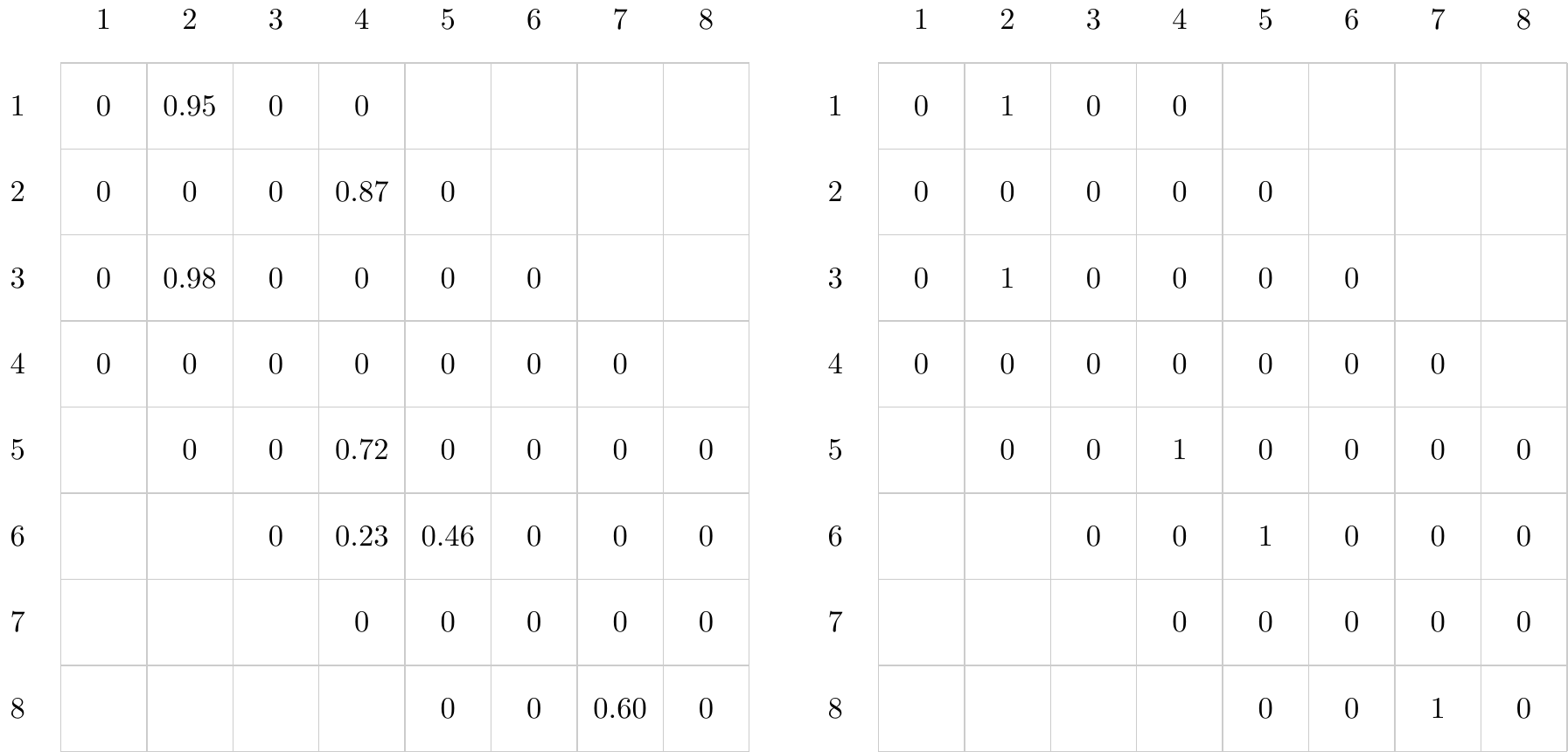}
    \caption[Partial pairing matrix]{Illustration of the procedure \texttt{MakeBinary()}, which takes a partial score matrix (left) as an input and produces a pairing matrix (right). We start with the highest alignment score and set the first two, 0.98 and 0.95 to 1. The third highest, 0.87, would result in bifurcation, so it is set to 0. The next three scores are set to 1, but the last non-zero entry, 0.23 will result in a barrel involving strands 4, 5, and 6, so it is set to 0. The resulting partial pairing matrix has three blocks, listed as a set of strands, (1,2,3), (4,5,6) and (7,8). Filling this matrix by either 0 or 1 would result in $2^{20}=1048576$ different matrices, but the restrictions placed on pairing matrices means there are only 97 valid completions.}
    \label{fig:partmat}
  \end{figure*}

  The above procedure allows us to populate $\mathbb{P}$ along the super- and sub-diagonals. We now repeat the procedure with $s$ being a segment containing $k$ $\beta$-strands, $k=1,2,3,\dotsc$, such that $s$ is the segment between $i$'th and $i+k+1$'th $\beta$-strands. We do this to populate $\mathbb{P}$ up to $d$ entries above and below the diagonal, where $d$ is given by;
  \begin{equation*}
    d =
    \begin{cases}
      1 & \text{ if } n<7\\
      n-5 & \text{ if } 7 \leq n < 11\\
      5 & \text{ if } 11 \leq n.
    \end{cases}
  \end{equation*}
  Here the limit of 5 for $d$ is forced by the fact that as the segments get longer, it becomes increasingly harder to obtain high alignment scores. This results in the  chance of having $\mathbb{P}_{(i,j)}=1$, in the discretisation process desribed below, being extremely small, when $\abs{i-j}>4$ (We were not able to get 1 in these cells in our tests). This is possibly related to the fact that the above method is essentially a method based on local data, and thus is not suited for predicting non-local configuration of $\beta$-strands. For that, another approach is needed which takes into account the global characteristics, which we will describe in the second part of the method. Before that, we need to translate the entries of the partial pairing matrix computed above, which are real numbers between 0 and 1, to either 0 or 1. We do this by changing the non-zero entries to 1, starting from the largest to the smallest. If, at any point, setting an entry to 1 results in a bifurcation or a $\beta$-barrel, the entry is set to 0 and we move onto the next largest entry (\Cref{fig:partmat}). For later use, we name this procedure \texttt{MakeBinary()}, which takes a (partial) matrix of pairing scores as an input and returns a (partial) pairing matrix.

  We now have a partial pairing matrix, populated up to $d$ entries above and below the diagonal, without bifurcations or barrels. We populate the remaining entries by going through all possibilities, while avoiding bifurcations and $\beta$-barrels. We also require that the resulting matrix does not contain any isolated strand. The result is a number of candidate matrices, whose number depends on the partial pairing matrix computed in the first part of the method. We now construct a fatgraph from each candidate matrix, and compute its genus and number of boundary components, together with the number of strands in the largest sheet. We compare this data with the 3-dimensional genus-boundary distribution computed in \Cref{sec:topochar}. By a layer in the 3-dimensional genus-boundary distribution, we mean the 2-dimensional distribution of genus and number of boundary components for a specific value of number of strands in the largest sheet. Let $g$, $n$, $l$ denote the genus, the number of boundary components and the number of strands in the largest sheet. Let $f(g,n,l)$ be the frequency of the cell $(g,n,l)$ in the 3-dimensional genus-boundary distribution. We define the topology score $s_{\mathrm{topo}}(\tau)$ for a metastructure $\tau$ with genus $g$, $n$ boundary components and $l$ strands in the largest sheet, by
  \begin{equation*}
    s_{\mathrm{topo}}(\tau) = \frac{f(g,n,l)}{T_l},
  \end{equation*}
where $T_l$ is the sum of frequencies for the $l$th layer. For a cutoff value $v \in (0,1)$, a candidate metastructure $\tau$ is accepted, if $s_{\mathrm{topo}}(\tau) \geq v$, and rejected if $s_{\mathrm{topo}}(\tau) < v$. We also compute accuracy of each candidate structure, and look at the relationship between accuracy and acceptance of candidate structures.

\subsection{Metastructure prediction by sequence alignment and topology}\label{sec:metastr_align}
The method described in \Cref{sec:binclas_metastr} was modified to provide a single, ``best candidate'' metastructure. The modification was made such that instead of classifying candidate metastructures as either accepted or rejected, a weighted sum of all candidate pairing matrices was produced, with weight given by the 3-dimensional genus-boundary distribution. More precisely, suppose a candidate pairing matrix $\mathbb{P}$ results in a structure with genus $g$, $n$ boundary components and $l$ strands in the largest sheet. Let $f(g,n,l)$ be the frequency of the cell $(g,n,l)$ in the 3-dimensional distribution, and $T_l$ be the sum of frequencies for the $l$th layer, as before. Then our final score matrix $\hat{\mathbb{P}}_\mathrm{score}$ is given by
\begin{equation*}
  \hat{\mathbb{P}}_\mathrm{score} = \sum_\mathbb{P} \frac{f(g,n,l)}{T_l} \mathbb{P},
\end{equation*}
where the sum is over all candidate pairing matrices for a protein. The final pairing matrix $\hat{\mathbb{P}}$ is computed from $\hat{\mathbb{P}}_\mathrm{score}$ as before. A pseudocode for this procedure is shown in \Cref{alg:pred_pm}.

\begin{algorithm}
  \begin{algorithmic}
    \State Let $\mathbb{P}_{\mathrm{partial}}$ be a given partial pairing matrix.
    \State Let $\hat{\mathbb{P}}_{\mathrm{score}}$ be a zero matrix of the same size as $\mathbb{P}_{\mathrm{partial}}$.
    \ForAll{Completion $\mathbb{P}$ of $\mathbb{P}_{\mathrm{partial}}$}
    \If{$\mathbb{P}$ contains a barrel, a bifurcation or an isolated strand}
    \State Move to next completion
    \EndIf
    \State Compute genus $g$, number of boundary components $n$, and size of the largest sheet $l$ for the metastructure corresponding to $\mathbb{P}$.
    \State Find the frequency of the cell $(g,n,l)$ and the sum of frequencies for the $l$th layer $T_l$.
    \State $\hat{\mathbb{P}}_{\mathrm{score}} = \hat{\mathbb{P}}_{\mathrm{score}} + \frac{f(g,n,l)}{T_l}\mathbb{P}$
    \EndFor
    \State Set $\hat{\mathbb{P}}$ = \Call{MakeBinary}{$\hat{\mathbb{P}}_{\mathrm{score}}$}
  \end{algorithmic}
  \caption{Pseudocode for computation of prediction pairing matrix $\hat{\mathbb{P}}$.}
  \label{alg:pred_pm}
\end{algorithm}

\subsection{Metastructure prediction by Betapro and topology}\label{sec:metastr_bp}
Betapro is a computer program for predicting $\beta$-sheet topology using recurrent neural network (RNN) \cite{cheng05}. It takes a primary structure sequence as input, or a primary and secondary structure sequences, if the secondary structure is available from other sources. The output is a score matrix, where the entries are not restricted to $(0,1)$, but positive real numbers computed as a sum of pseudoenergy for each residue pair in a $\beta$-strand pairing. The reported performances of Betapro are 0.54 for Recall and 0.59 for Precision \cite{cheng05}.

In order to predict protein metastructure, we run Betapro using the primary and secondary structure sequences as input. From the output score matrix, we choose $m$ entries with the highest scores, where $m$ equals 4\% of the number of entries in the score matrix, excluding the main diagonal. The entries that result in a bifurcation or a barrel, are ignored. The chosen entries are considered as $\beta$-strand pairings, and they are set to 1 in the partial pairing matrix. Next, all valid (i.e. avoiding isolated strands, bifurcations and barrels) completions of the partial pairing matrix are generated. Each completion is given two scores, one based on Betapro score matrix, and the other based on the genus-boundary distribution. The first, $s_\mathrm{bp}$, is the sum of all scores in Betapro score matrix, where there is 1 in the pairing matrix. The second, $s_\mathrm{topo}$, is given by $f(g,n,l)/T_l$, where $g, n, l$ is the genus, the number of boundary components, and size of the largest sheet, as before. Our prediction is the structure with the highest combined score,
\begin{equation}\label{eq:metastr_score}
  \hat{s} = a s_\mathrm{bp} + b s_\mathrm{topo},
\end{equation}
where $a, b \in [0,1]$ with $a+b=1$. The corresponding pseudocode is shown in \Cref{alg:pred_bp}.

\begin{algorithm}
  \begin{algorithmic}
    \State Let $\mathbb{P}_{\mathrm{bp}}$ be the pairing score matrix produced by Betapro.
    \State Let $\mathbb{P}_{\mathrm{partial}}$ be an empty matrix of the same size as $\mathbb{P}_{\mathrm{bp}}$.
    \State Order the entries in $\mathbb{P}_{\mathrm{bp}}$ from largest to smallest.
    \State Set $c=0$.
    \While {$c \leq m$}
    \State Let $(i,j)$ be the index for the first element in the ordered list of entries in $\mathbb{P}_{\mathrm{bp}}$.
    \State Set $\mathbb{P}_{\mathrm{partial}(i,j)} = 1$.
    \If {$\mathbb{P}_{\mathrm{partial}}$ results in a barrel or a bifurcation}
    \State Set $\mathbb{P}_{\mathrm{partial}(i,j)} = 0$.
    \State $c = c-1$.
    \EndIf
    \State Remove the first element from the ordered list of entries in $\mathbb{P}_{\mathrm{bp}}$.
    \State $c = c+1$
    \EndWhile
    \ForAll {Completion $\mathbb{P}$ of $\mathbb{P}_{\mathrm{partial}}$}
    \If{$\mathbb{P}$ contains a barrel, a bifurcation or an isolated strand}
    \State Move to next completion
    \EndIf
    \State Compute genus $g$, number of boundary components $n$, and size of the largest sheet $l$ for the metastructure corresponding to $\mathbb{P}$.
    \State Find the frequency of the cell $(g,n,l)$ and the sum of frequencies for the $l$th layer $T_l$.
    \State Set $s_{\mathrm{topo}}(\mathbb{P}) = \frac{f(g,n,l)}{T_l}$.
    \State Set $\mathbb{P}_{\mathrm{score}} = \mathbb{P} \dot{\times} \mathbb{P}_{\mathrm{bp}}$, where $\dot{\times}$ denotes the entry-wise multiplication.
    \State Set $s_{\mathrm{bp}}(\mathbb{P}) =  \sum_{i,j} \mathbb{P}_{\mathrm{score}(i,j)}$.
    \State Set $\hat{s}(\mathbb{P}) = a s_{\mathrm{bp}}(\mathbb{P}) + b s_{\mathrm{topo}}(\mathbb{P})$.
    \EndFor
    \State Set $\hat{\mathbb{P}}$ to be the completion $\mathbb{P}'$, such that $\hat{s}(\mathbb{P}') = \max\{\hat{s}(\mathbb{P}) | \mathbb{P} \text{ is a completion of } \mathbb{P}_{\mathrm{partial}}\}$.
  \end{algorithmic}
  \caption{Pseudocode for computation of prediction pairing matrix $\hat{\mathbb{P}}$ from Betapro score matrix $\mathbb{P}_{\mathrm{bp}}$.}
  \label{alg:pred_bp}
\end{algorithm}

\section{Results}
\begin{figure}[h]
  \centering
  \begin{subfigure}[b]{\linewidth}
    \includegraphics[width=.9\linewidth]{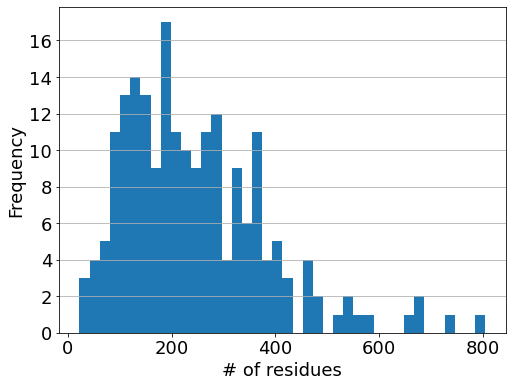}
    \caption{By the number of residues}
  \end{subfigure}
  \begin{subfigure}[b]{\linewidth}
    \includegraphics[width=.9\linewidth]{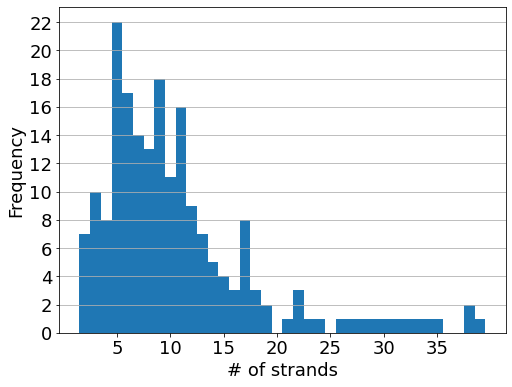}
    \caption{By the number of strands}
  \end{subfigure}
  \caption[Distribution of proteins]{Frequency distribution of 200 proteins by the number residues (a) and by the number of strands (b).}
  \label{fig:prot_freq}
\end{figure}

Some of the larger proteins in the 200 test proteins could not be analysed using the method described, as there were too many possible ways to complete the pairing matrix. We therefore limit the analysis to the 181 proteins containing up to 20 $\beta$-strands. Their frequency distribution by the number of residues and $\beta$-strands is shown in \Cref{fig:prot_freq}.

\begin{figure}[h]
  \centering
  \includegraphics[width=.9\linewidth]{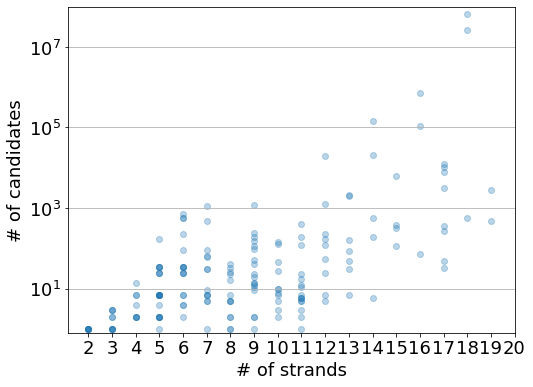}
  \caption[Number of candidates]{Number of candidate structures per protein, filtered by the number of strands. Note the log scale. There are large variations in the number of candidates among the proteins with the same number of strands.}
  \label{fig:num_cands}
\end{figure}

The algorithm from \Cref{sec:binclas_metastr} produced 91,431,292 candidate structures in total, but there are significant variations in the number of candidate structures per protein (\Cref{fig:num_cands}), as the possible number of candidates also depends on the partial structure determined using alignment of the $\alpha/\gamma$ segments between $\beta$-strands. In the current analysis, one protein (4UPIA) accounted for 63,907,920 candidate structures, representing 70\% of the total number. Note, although some of these numbers are large, they still represent a significant reduction from the theoretically possible number of candidate structures, which is given by $n! \cdot 2^{n-2}$ for a protein with $n$ strands, when considering only those structures with a single sheet. Naturally the numbers are even larger when considering multiple-sheet structures. We list the first few terms in \Cref{tab:num_str}. 

\begin{table}[h]
  \centering
  \begin{tabular}[c]{lll}
    \multirow{2}{*}{Strands}
    & \multicolumn{2}{c}{Number of structures} \\
    & Single sheet & Multiple sheets \\
    \hline
    2 & 2 & 2 \\
    3 & 12 & 12 \\
    4 & 96 & 108 \\
    5 & 960 & 1200 \\
    6 & 11520 & 15960 \\
    7 & 161280 & 246960 
  \end{tabular}
  \caption[Number of possible structures]{The number of theoretically possible structures for a protein with $n$ strands.}
  \label{tab:num_str}
\end{table}

\begin{figure}[h]
  \centering
  \includegraphics[width=.9\linewidth]{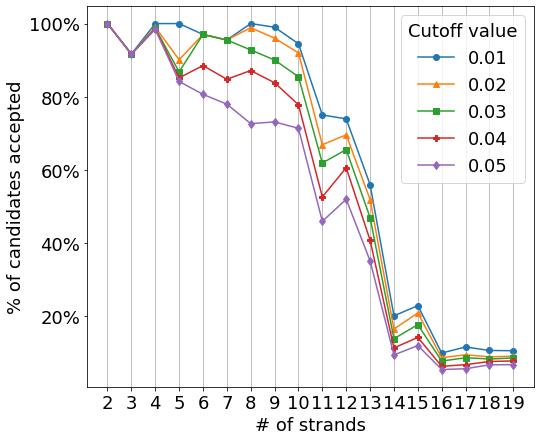}
  \caption[Acceptance rate for candidate structures]{Percentages of accepted structures by cutoff values and the number of strands.}
  \label{fig:accept}
\end{figure}

\begin{figure}[h]
  \centering
  \includegraphics[width=.9\linewidth]{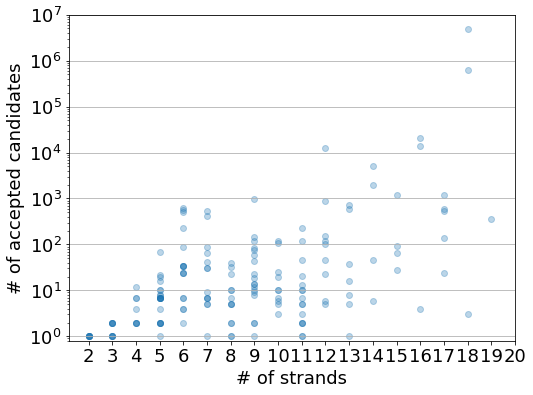}
  \caption[Number of accepted candidates]{Number of accepted candidate structures per protein, filtered by the number of strands. Note the log scale. Compared to \Cref{fig:num_cands}, the numbers are significantly lower where there are large number ($>10^4$) of candidate structures.}
  \label{fig:num_acc}
\end{figure}

The topology filter, depending on the cutoff value and the number of strands, further reduces the number of candidate structures (\Cref{fig:accept}). Upon considering the balance between the ability to reduce the number of candidate structures and still retain high quality candidate structures, we decided to use the cutoff $s_{\mathrm{topo}}$ value of 0.02 for the subsequent analysis. The actual number of accepted candidate structures are shown in \Cref{fig:num_acc}. As we also can see from \Cref{fig:accept}, the topology filter is very effective at reducing the number of structures for proteins with larger number of strands (i.e. large number of candidates). In the current analysis, 3 proteins accounted for 99.6\% of all candidate structures. For these the topology filter reduced the number of candidates by 92-97\% (\Cref{tab:reduction}). When using any positive cutoff value for such a filter, there is a chance that no candidate structure for a protein is accepted. If it happens, we reduce the cutoff value only for the proteins with no accepted candidate structure, until one or more candidate structures are accepted. In the current analysis, the cutoff values were reduced by 0.005 down to 0.005. If, at the end of this iteration, we have proteins with no accepted structure, we randomly select one candidate structure for acceptance. This procedure, however, was not necessary for the current analysis, and all proteins had at least one candidate structure accepted at the cutoff value of 0.02.


\begin{table}[h]
  \centering
  \begin{tabular}[c]{rrr}
    \# candidates & \# accepted & \% accepted \\
    \hline
    63907920 & 4889934 & 7.7\%\\
    26437952 & 631882  & 2.4\%\\
    731584  & 20521  & 2.8\%
  \end{tabular}
  \caption{The numbers and percentages of accepted structures for the three proteins with most candidate structures, accounting for more than 99\% of all candidates. The topology filter rejects more than 90\% of candidates.}
  \label{tab:reduction}
\end{table}

In order to examine how well our topological filter distinguishes between ``good'' and ``bad'' candidates, we investigate how the rate of acceptance changes for ``good'' and ``bad'' candidate structures. Precision and Recall are two measures often used for judging quality of predicted protein structures. They are given by
\begin{align*}
  \mathrm{Precision} &= \frac{\mathrm{TP}}{\mathrm{TP} + \mathrm{FP}}\\
  \mathrm{Recall} &= \frac{\mathrm{TP}}{\mathrm{TP} + \mathrm{FN}},
\end{align*}
where $\mathrm{TP}, \mathrm{FP}$ and $\mathrm{FN}$ stand for the number of true positive, false positive and false negative strand pairings.

For each target and for a given quality measure $Q$ (=Precision or Recall), we divide the candidate structures into three classes; low quality (structures with $Q<0.6$), medium quality ($0.6 \leq Q < 0.9$), and high quality ($0.9 \leq Q$). We then compute the acceptance rate for each class. The results are shown in \Cref{tab:acc_rate}. The acceptance rates increase with an increase in the quality levels.

\begin{table}[h]
  \centering
  \begin{tabular}[c]{l|rr}
    Quality & Precision & Recall \\
    \hline
    Low & 66.33\% & 65.45\% \\
    Medium & 70.71\%  & 72.04\% \\
    High & 90.80\%  & 89.62\% \\
  \end{tabular}
  \caption[Acceptance rates]{Average acceptance rate by quality classes.}
  \label{tab:acc_rate}
\end{table}

\begin{figure}[h]

  \begin{subfigure}[b]{0.5\linewidth}
    \includegraphics[scale=.35]{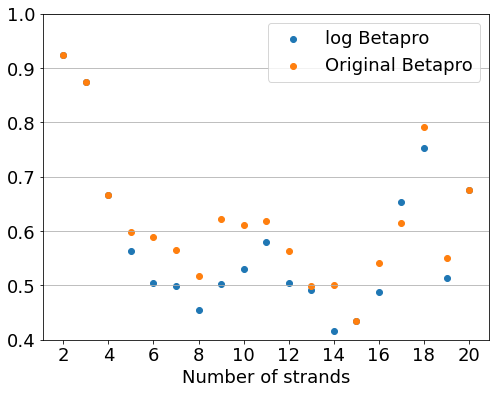}
    \caption{Average Recall}
  \end{subfigure}
  \begin{subfigure}[b]{0.5\linewidth}
    \includegraphics[scale=.35]{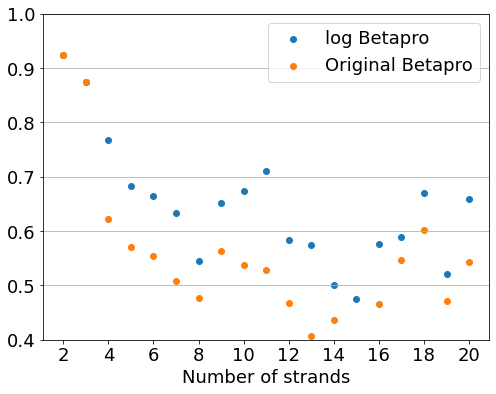}
    \caption{Average Precision}
  \end{subfigure}
  \caption[logBetapro, Recall and Precision]{Average Recall (a) and Precision (b) by number of strands, Betapro and logBetapro scores.}
  \label{fig:logbp}
\end{figure}

Metastructure prediction by sequence alignment and topology (\Cref{sec:metastr_align}) and by Betapro and topology (\Cref{sec:metastr_bp}) were performed on the same set of proteins. The average Precision and Recall for the predictions are shown in \Cref{tab:metastr_pred}. Different values of $a$ in the combined score function \eqref{eq:metastr_score} only had a very small effect ($<0.005$) on Precision or Recall values (\Cref{tab:metastr_pred}). The strand pairing scores from Betapro are strictly positive, potentially promoting the formation of large sheets which are topologically complex. To mitigate this, we applied logarithm to the strand pairing scores from Betapro and used them in the algorithm. This resulted in an increase in Precision but a (smaller) decline in Recall (\Cref{tab:metastr_pred}). This change was seen across different number of strands (\Cref{fig:logbp}). To investigate the effect of the number of selected pairings before computing completions, we ran the algorithm using 4, 5, and 6\% for pre-selection, together with the ``fewest possible'' pre-selections, which is the number where a computation is possible within a reasonable amount of time (24 hours on a modern cpu). The number $p$ of pre-selected pairs for a protein with $n$ strands was;
\begin{align*}
  p =
  \begin{cases}
  0 \qquad &\text{ if } n \leq 8 \\
  n-8 \qquad &\text{ if } 9 \leq n \leq 11 \\
  n-7 \qquad &\text{ if } 12 \leq n \leq 20    
  \end{cases}
\end{align*}
The results are shown in \Cref{tab:preselection}. Pre-selecting more pairings should have the effect of increasing false positive (FP) and decreasing false negative (FN), thereby reducing Precision and increasing Recall, which we observe here.

\begin{table}[h]
  \centering
  \begin{tabular}[c]{l|rr}
    & Precision & Recall \\
    \hline
    Alignment & 0.42 & 0.47 \\
    Betapro, a=0.1 & 0.56 & 0.62 \\
    Betapro, a = 1 & 0.56 & 0.62 \\
    logBetapro & 0.67 & 0.57 \\
  \end{tabular}
  \caption[Metastructure prediction results]{Average Precision and Recall for different metastructure prediction methods.}
  \label{tab:metastr_pred}
\end{table}

\begin{table}[h]
  \centering
  \begin{tabular}[c]{l|rrrr}
    & Fewest & 4\% & 5\% & 6\% \\
    \hline
    Precision & 0.667 & 0.667 & 0.659 & 0.650 \\
    Recall & 0.567 & 0.567 & 0.572 & 0.577 \\
  \end{tabular}
  \caption[Recall and Precision by pre-selection levels]{Average Precision and Recall for different levels of pre-selection.}
  \label{tab:preselection}
\end{table}

\section{Discussion}
The difference in the distributions of the genera and the numbers of boundary components from the actual (\Cref{fig:heatmap_pdb}) and simulated data (\Cref{fig:heatmap_sim}) indicate that the folding of $\beta$-sheets is not a completely random process. Indeed, it does appear that an increase in genus is costly and a structure that has lower genus is favoured over one with higher genus. This observation agrees with previous studies, which do not look at genus of $\beta$-sheets, but finds that certain $\beta$-sheet structures, many of which correspond to an increase in genus, are absent or very rare in proteins \cites{ruczinski02, zhang00}. The result of our binary classification analysis (\Cref{sec:binclas_metastr}) agrees with this observation. Even though the result is skewed by a highly uneven distribution of the number of candidate structures per protein, and the response of acceptance rate for an increase in quality is not linear, it does appear that the topology of protein metastructure captures some information about the native structure. Extending this result to prediction of metastructures proved more challenging. We did achieve a result comparable to that reported for Betapro when using strand-pairing scores as is, which was improved to be better than Betapro with an application of logarithm to the pairing scores. This is likely to be because the unprocessed Betapro scores are strictly greater than zero, thus encouraging formation of larger sheets in order to maximise the final score $\hat{s}$, even though the contribution from the topology score $s_{\mathrm{topo}}$ should, to some extent, prevent the formation of sheets that are too large and topologically complex. By applying logarithm to the Betapro scores, we encourage fewer pairings (and thus discourage large sheets), which resulted in improved Precision. We were, however, not able to outperform the figures reported by other, more recent studies (\Cref{tab:metastr_comp}). The structure of the BCov and BetaProbe programs meant that it was not possible to combine them with our method in a similar manner to \Cref{sec:metastr_bp}. It would be interesting to see if one can improve the results of Top-DBS program by combining with our method. Unfortunately the program code for Top-DBS was not available for inspection.

\begin{table}[h]
  \centering
  \begin{tabular}[c]{crr}
    Program & \multicolumn{1}{c}{Precision} & \multicolumn{1}{c}{Recall} \\
    \hline
    Betapro \cite{cheng05} & 0.59 & 0.54 \\
    BCov \cite{savojardo13} & 0.60 & 0.62 \\
    BetaProbe \cite{eghdami15} & 0.67 & 0.70 \\
    Top-DBS \cite{dehghani18} & 0.75 & 0.78 \\
    Current Study & 0.67 & 0.57
  \end{tabular}
  \caption[Metastructure prediction comparison]{Comparison of Precision and Recall values for prediction of $\beta$-sheet topology.}
  \label{tab:metastr_comp}
\end{table}

One of the reasons why the results from our study could not match those from more recent studies may be that the topology filter, in its current form, is too coarse. Suppose we have a protein with three $\beta$-strands. There are 12 different protein metastructure configurations possible, but 8 of them have genus 0 and 3 boundary components, with the rest having genus 1 and 1 boundary component. This suggests a ``finer'' filter, which can distinguish between the structures having the same genus and number of boundary components (and maximum sheet size), may be able to produce a better result. However, with the size of the currently available dataset, making the filter finer would result in the frequency in each cell being too small for sampling the distribution of genera or numbers of boundary components (or some other topological data).

The term $\beta$-sheet topology is commonly used to describe the configuration of $\beta$-strands in a $\beta$-sheet. However, to our knowledge, it has not been studied in relation to topological invariants. We have shown in this paper that the topological invariants such as genus and the number of boundary components can describe certain aspects of $\beta$-sheet topology of proteins, and how they might be used in prediction of $\beta$-sheet topologies. We believe the protein metastructure and topology of the associated fatgraph have a potential to provide a simpler, more mathematically natural way to analyse $\beta$-sheet topology.

\subsubsection*{Acknowledgement}
This paper is partly a result of the ERC-SyG project, Recursive and Exact New Quantum Theory (ReNewQuantum) which received funding from the European Research Council (ERC) under the European Union's Horizon 2020 research and innovation programme under grant agreement No. 810573. This work was supported by JSPS KAKENHI Grant Number JP20K03931, JP20K03601, JP18K03281.

\printbibliography

\end{document}


\beginsupplement
\maketitle

\section{Extension of Blosum62}\label{app:blosum}
\setcounter{figure}{0}

\begin{figure}[h]
  \centering
  \begin{subfigure}[b]{.45\linewidth}
    \includegraphics[scale=.3]{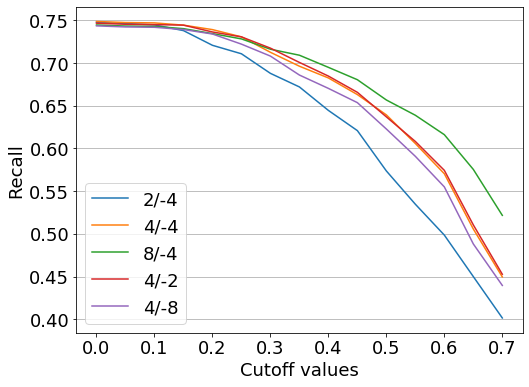}
    \caption{Average Recall}
    \label{fig:blosum_rec}
\end{subfigure}
\begin{subfigure}[b]{.45\linewidth}
  \includegraphics[scale=.3]{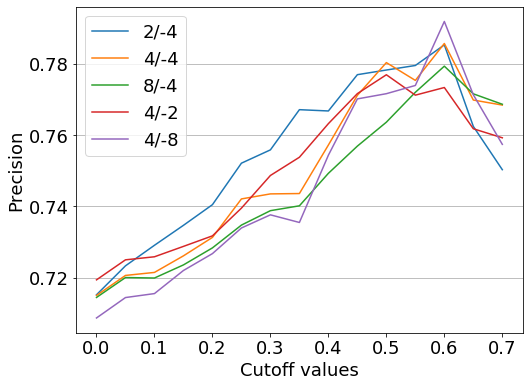}
  \caption{Average Precision}
  \label{fig:blosum_prec}
\end{subfigure}
  \caption[]{Average Recall (\Cref{fig:blosum_rec}) and Precision (\Cref{fig:blosum_prec}) for various cutoff values. The different lines represent different match/mismatch score combinations.}
  \label{fig:blosum}
\end{figure}

The sequences we align consists of 20 letters representing standard gene code amino acids and two extra letters, $\alpha$ and $\beta$. We therefore need to extend the blosum62 substitution matrix to include scores for these two extra letters. We use 4 and -4 respectively for match and mismatch involving $\alpha$ and $\beta$. We investigated the effect of these scores by computing average Recall and Precision for different scores as follows;
\begin{enumerate}
\item For each match/mismatch score combination, compute alignment scores for the first and second diagonal (i.e. for the sequences involving zero or one $\beta$-strand).
\item Let $v$ be a number between 0 and 1. For each cell $\mathbb{P}_{(i,j)}$ in the pairing matrix $\mathbb{P}$, where the alignment scores have been computed, set the value to 1 if $\mathbb{P}_{(i,j)}>v$, 0 otherwise.
  \item Compute Recall and Precision for each protein.
  \end{enumerate}
The average Recall and Precision for various cutoff values and score combinations are shown in \Cref{fig:blosum}. We see the variation in Precision is very small across different score combinations. The same holds for Recall, for small cutoff values. We also note that, compared to the average Recall, the average Precision does not vary much for different cutoff values. We therefore choose the cutoff value of 0 and match/mismatch score combination of 4 and -4 for the current analysis.

\clearpage
\section{The topology filter}\label{app:filter}

\begin{figure}[h]
  \centering
  \includegraphics[width=\linewidth]{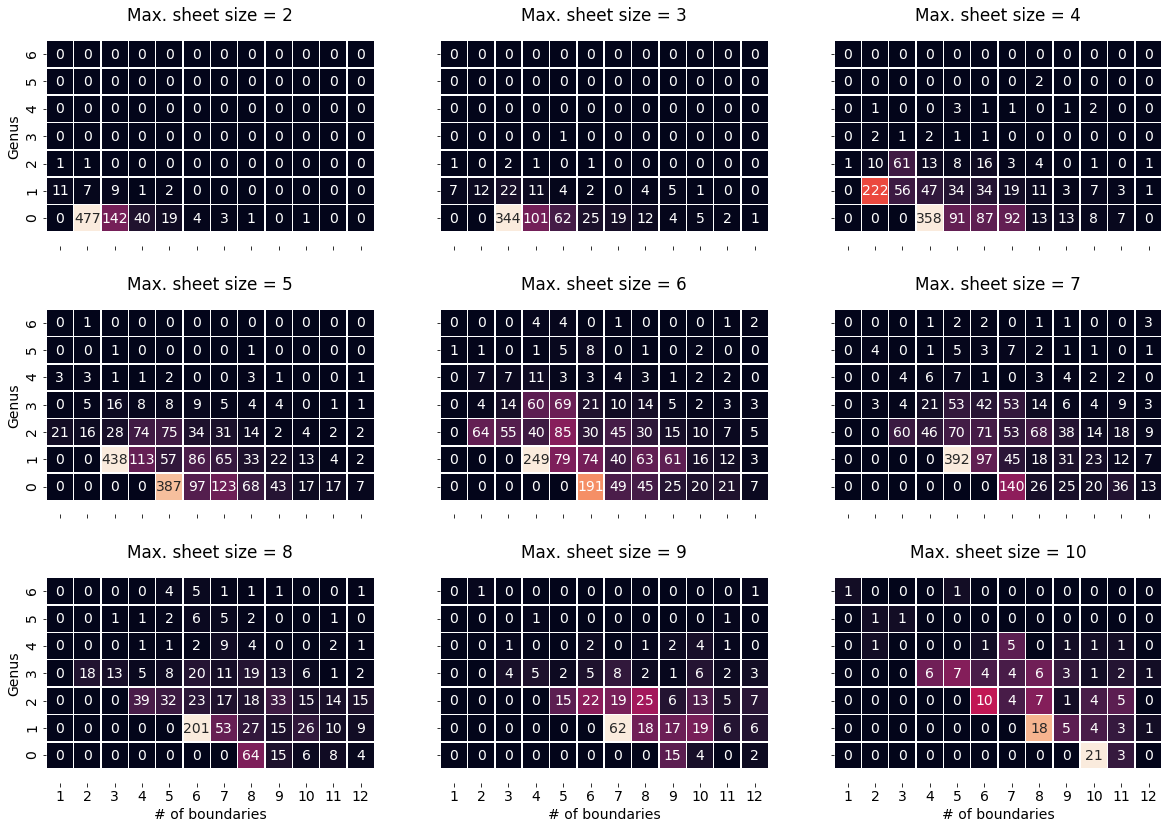}
  
  \caption{The distribution of genus and number of boundary components, filtered by the size of the largest sheet.}
  \label{fig:dist}
\end{figure}

\begin{figure}[h]
  \centering
  \includegraphics[width=\linewidth]{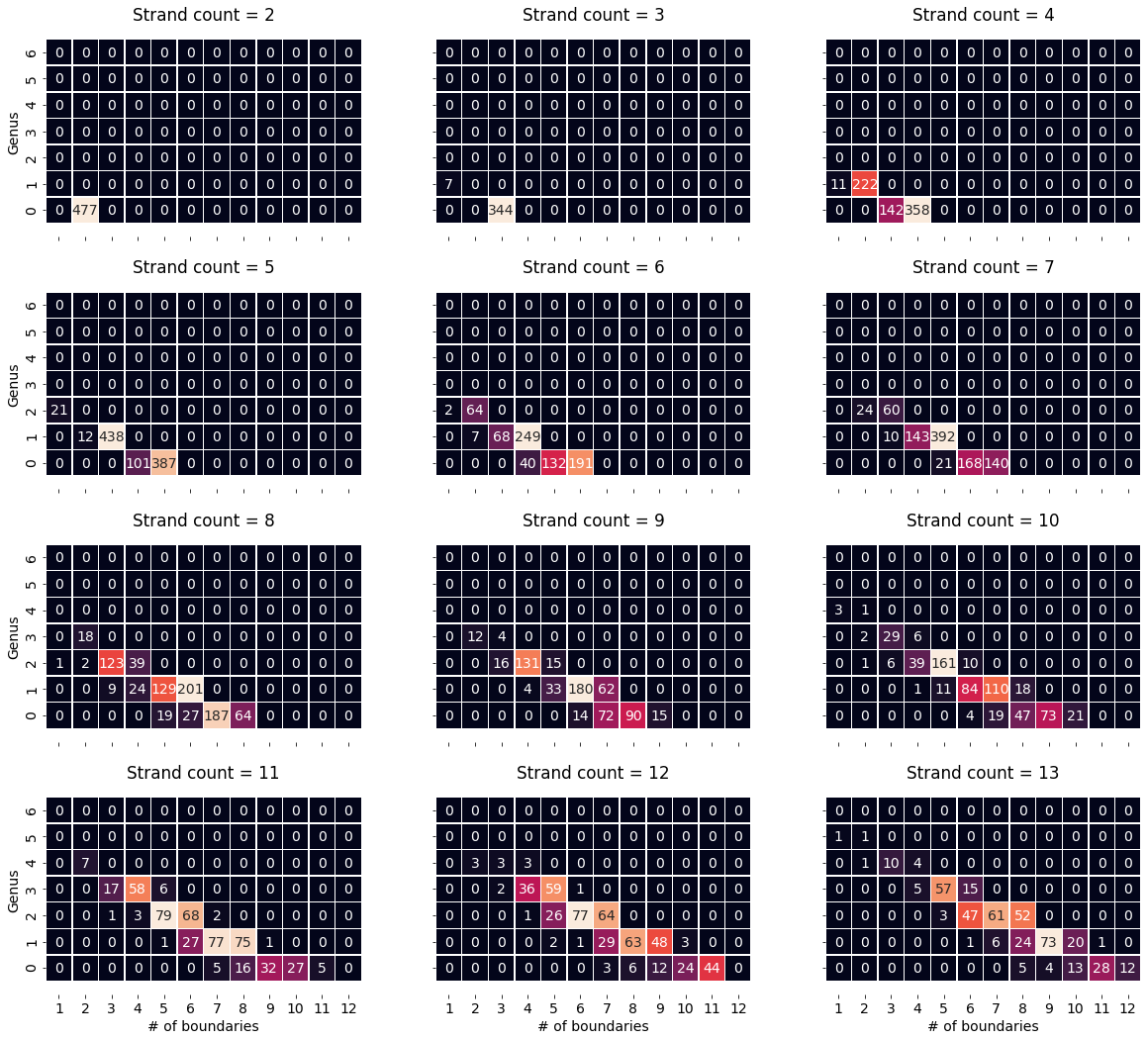}
  
  \caption{The distribution of genus and number of boundary components, filtered by the number of strands.}
  \label{fig:dist_strds}
\end{figure}

\begin{figure}[h]
  \centering
  \includegraphics[width=\linewidth]{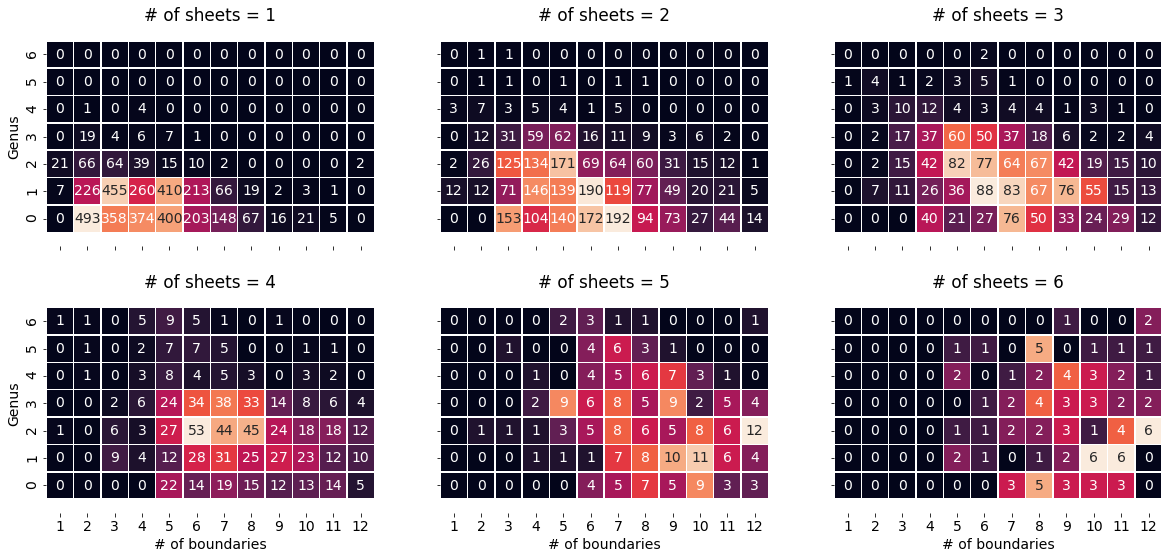}
  
  \caption{The distribution of genus and number of boundary components, filtered by the number of sheets.}
  \label{fig:dist_sheets}
\end{figure}

\begin{figure}[h]
  \centering
  \begin{subfigure}[b]{.45\linewidth}
    \includegraphics[scale=.3]{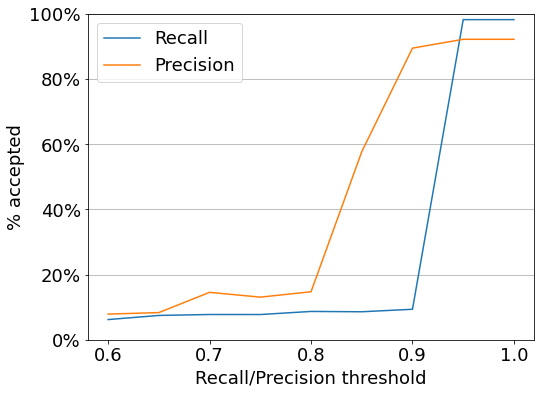}
    \caption{Maximum sheet size}
    \label{fig:pc_maxsheet}
  \end{subfigure}
  \quad
\begin{subfigure}[b]{.45\linewidth}
  \includegraphics[scale=.3]{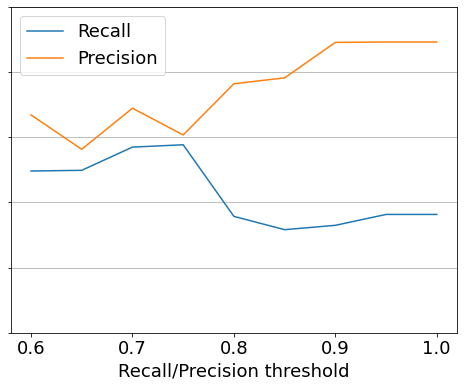}
  \caption{Number of strands}
  \label{fig:pc_nstrands}
\end{subfigure}
  \caption[Acceptance rates for different filters]{Acceptance rates for candidates above quality thresholds (measured in Recall and Precision). The topology filter with maximum sheet size (\Cref{fig:pc_maxsheet}) shows increasing acceptance rates when the candidates are restricted to high-quality structures. On the other hand, the filter with number of strands (\Cref{fig:pc_nstrands}) shows relatively high acceptance rates for lower-quality candidates, and the rate drops for Recall, when the candidates are restricted to high-quality structures.}
  \label{fig:pc_accept}
\end{figure}

The topology filter, as used in the paper, is the distribution of genera and numbers of boundary components for the protein metastructures filtered by the number of strands in the largest sheet. The filtering was done to be able to reflect the difference in the distributions between proteins with a small number of strands and those containing more strands, as one would expect those proteins with a large number of strands to have more (topologically) complex structures. The number of strands in the largest sheet was chosen as the filtering variable, because we expect the largest contribution to the genus (and the number of boundary components) to come from the largest sheet. We show the distributions up to the maximum sheet size of 10 in \Cref{fig:dist}. We also show the distributions of the same data, filtered by the number of strands \Cref{fig:dist_strds} and by the number of sheets \Cref{fig:dist_sheets}. It was thought that filtering by the number of sheets results in too few layers and will make the resulting topology filter less powerful, as it will not be able to distinguish subtler differences. To investigate whether filtering by the number of strands produces better results, we ran the binary classification (see Section 3.1) using the topology filter with the maximum sheet size as the third axis, and one with the number of strands as the third axis. The classification results were analysed by computing the proportion of the candidate structures above certain quality thresholds, that were accepted (\Cref{fig:pc_accept}). For a good filter, we expect the percentages of acceptance to increase, as we restrict to candidate structures to only look at the high-quality structures. So we expect the lines to lie diagonally from the bottom-left to top-right. It was found the filter using the maximum sheet size performed better, particularly with Recall. We did not investigate why it is the case, but it may be that folding several large sheets into energetically favourable structure is complex, and in nature a combination of one large sheet and several smaller ones is more common. 

\clearpage
\printbibliography